\documentclass{PoS}
\usepackage{amsopn}
\usepackage{graphicx}
\graphicspath{{./plots/}}

\newcommand{\lr}[1]{ \left( #1 \right) }
\newcommand{\lrs}[1]{ \left[ #1 \right] }

\newcommand{\vev}[1]{ \langle \, #1 \, \rangle }

\newcommand{\bra}[1]{ \langle #1 | }
\newcommand{\ket}[1]{ | #1 \rangle }

\DeclareMathOperator{\sign}{sign}

\title{First experience with classical-statistical real-time simulations of anomalous transport with overlap fermions}

\ShortTitle{Real-time simulations with overlap fermions}

\author{\speaker{S.~N.~Valgushev}\\
        Institut f\"ur Theoretische Physik, Universit\"at Regensburg, 93053 Regensburg, Germany\\
        E-mail: \email{semen.valgushev@ur.de}}

\author{P.~V.~Buividovich\thanks{This work was supported by the S.~Kowalevskaja award from the Alexander von Humboldt foundation.}\\
        Institut f\"ur Theoretische Physik, Universit\"at Regensburg, 93053 Regensburg, Germany\\
        E-mail: \email{pavel.buividovich@physik.uni-regensburg.de}}

\abstract{We present first results of classical-statistical real-time simulations of anomalous transport phenomena with overlap fermions. We find that even on small lattices overlap fermions reproduce the real-time anomaly equation with much better precision than Wilson-Dirac fermions on an order of magnitude larger lattices. The difference becomes much more pronounced for quickly changing electromagnetic fields, especially if one takes into account the back-reaction of fermions on electromagnetism. As test cases, we consider chirality pumping in parallel electric and magnetic fields and mixing between the plasmon and the Chiral Magnetic Wave.}

\FullConference{34th annual International Symposium on Lattice Field Theory\\
		24-30 July 2016\\
		University of Southampton, UK}

\begin{document}
\sloppy

\section{Introduction}
\label{sec:intro}

Transport phenomena related to quantum anomalies of chiral fermions might have profound impact on the properties of dense chiral medium, which can be realized in Dirac or Weyl semimetals, in neutrino/leptonic matter in astrophysical contexts and in quark-gluon plasma. It is by now commonly accepted that such phenomena as the Chiral Magnetic Effect and the Chiral Magnetic Wave cannot exist in equilibrium \cite{Yamamoto:15:1}. For this reason first-principle numerical studies of anomalous transport require real-time simulation techniques, especially if the relevant dynamics is strongly nonlinear, as in the case of chiral plasma instability \cite{Yamamoto:13:1,Buividovich:15:2} or chiral shock waves \cite{Yamamoto:16:2}.

Currently the state-of-the-art method for real-time simulations is the classical-statistical approximation, in which the gauge fields are treated as classical, but the dynamics of fermions is fully quantum. So far most real-time simulations of anomalous transport phenomena have been performed with Wilson-Dirac fermions, for which the chiral symmetry is explicitly broken by the Wilson term. At the same time, Wilson term is responsible for axial anomaly of Wilson fermions (see e.g. \cite{Berges:16:1} for a recent real-time study). It is therefore important to understand, how lattice artifacts of Wilson fermions can affect anomalous transport, and how these artifacts can be reduced. One practical solution considered recently in \cite{Mueller:16:1} is the improvement of the Wilson-Dirac operator which removes lattice artifacts up to some fixed order $\mathcal{O}\lr{a^n}$ in lattice spacing $a$.

In these Proceedings we advocate the use of overlap Hamiltonian, first proposed in \cite{Creutz:01:1}, for real-time classical-statistical simulations of chiral medium. We demonstrate that even on small lattices the real-time dynamics of overlap fermions reproduces the results known for continuum Dirac fermions with very good precision. In contrast, for Wilson-Dirac fermions the effects of explicit chiral symmetry breaking is significant even on lattices as large as $200 \times 200 \times 200$, especially for quickly changing electromagnetic fields. As a test case, we consider the generation of axial charge density in parallel electric and magnetic fields. If gauge fields are considered dynamical, this process corresponds to the formation of Chiral Magnetic Wave mixed with plasmon.

\section{Chirality pumping, Chiral Magnetic Wave and plasmon}
\label{sec:CMW_in_magnetic_field}

Let us consider a neutral plasma of massless charged particles in the background of external magnetic field $\vec{B} = \vec{e}_3 B$ and a time-dependent electric field $\vec{E} = \vec{e}_3 E\lr{t, x_3}$ parallel to $\vec{B}$ (here $\vec{e}_3$ is the unit basis vector along the $x_3$ coordinate axis). The axial charge and current $q_A$ and $\vec{j}_A$ obey the real-time anomaly equation
\begin{eqnarray}
\label{eq:anomaly}
 \partial_t q_A\lr{t, x_3} + \partial_3 j_{A \, 3}\lr{t, x_3} = \kappa \vec{E} \cdot \vec{B} = \kappa B E\lr{t, x_3} ,
\end{eqnarray}
where $\kappa = 1/\lr{2 \pi^2}$ is the anomaly coefficient (we assume unit electric charge). If electromagnetic field is dynamical, the longitudinal electric field $E\lr{t, x_3}$ satisfies the Maxwell equations
\begin{eqnarray}
\label{eq:Maxwell_equation}
 \partial_3 E(t, x_3) = q(t, x_3),
 \quad
 \partial_t E(t, x_3) = - j_3(t, x_3) ,
\end{eqnarray}
where $q\lr{t, x_3}$ and $j_3\lr{t, x_3}$ are the electric charge and current densities. In sufficiently strong magnetic fields $\lr{3+1}$-dimensional Dirac fermions can be effectively described as $\lr{1+1}$-dimensional Dirac fermions on the lowest Landau level. Higher Landau levels remain decoupled as long as all the relevant scales are significantly smaller than the energy of the next Landau level $E_1 = v_F \sqrt{2 B}$, where $v_F$ is the Fermi velocity. As a consequence of such dimensional reduction, longitudinal components of axial and electric currents are related to the electric and axial charge densities as
\begin{eqnarray}
\label{eq:Dimesional_reduction}
 j_3(t, x_3) = v_F \, q_A(t, x_3), \quad j_{A \, 3}(t, x_3) = v_F \, q(t, x_3).
\end{eqnarray}
Combining these equations with the Maxwell equations (\ref{eq:Maxwell_equation}) and the anomaly equation (\ref{eq:anomaly}), we arrive at the wave equation $\partial_t^2 E(t, x_3) - v_F^2 \partial_3^2 E(t, x_3) + v_F \, \kappa B E(t, x_3) = 0$, which describes the Chiral Magnetic Wave (CMW) \cite{Kharzeev:11:1} of axial/vector charge density propagating along the magnetic field. If electromagnetic fields are dynamical, the mixing of CMW with plasmon leads to the dispersion relation with a finite gap $\omega_A$:
\begin{eqnarray}
\label{eq:cmw_dispersion}
 \omega\lr{k_3} = \pm \sqrt{v_F^2 k_3^2 + \omega_A^2}, \quad \omega_A = \sqrt{v_F \kappa B} .
\end{eqnarray}

Particularly instructive is the spatially homogeneous solution with $E\lr{t = 0} = E_0$ and $q_A\lr{t=0} = q\lr{t = 0} = j_3\lr{t = 0} = j_{A \, 3}\lr{t = 0} = 0$:
\begin{eqnarray}
\label{eq:Plasma_oscillations}
 E(t) = E_0 \cos(\omega_A t),
 \quad
 q_A(t) = v_F^{-1} j_3(t) = \omega_A/v_F \, E_0 \sin(\omega_A t)
 =
 E_0 \sqrt{\kappa B} \sin\lr{\sqrt{\kappa B} t} .
\end{eqnarray}
At $\omega_A t \ll 1$ this solution is close to the behavior of the axial charge density $q_A\lr{t} = \kappa E_0 B \, t$ in the absence of backreaction of fermionic current on electromagnetic field. However, at later times the effect of backreaction turns the unbounded growth of axial charge density into periodic oscillations of amplitude $\omega_A E_0 = \sqrt{\kappa B} E_0$. It is interesting that the scaling of the axial charge density $q_A$ with $B$ changes from $q_A \sim B$ to $q_A \sim \sqrt{B}$. Simplicity of the solution (\ref{eq:Plasma_oscillations}) and apparent connection to anomaly makes it ideal candidate for testing real-time anomaly in lattice classical-statistical simulations with dynamical gauge fields.

\section{Overlap fermions within the classical-statistical field theory approximation}
\label{sec:CSFT}

As described in detail in a number of papers \cite{Buividovich:15:2,Mueller:16:1,Tanji:13:1,Berges:14:1}, classical-statistical approximation for gauge theories with fermions amounts to solving the quantum Heisenberg equations for fermionic fields $\hat{\psi}_x$ in the background of the classical gauge vector potential $\vec{A}_x$, which in turn satisfies the Maxwell (or Yang-Mills) equations with the fermionic current term $\vec{j}_x$:
\begin{eqnarray}
\label{eq:csft_eqs_fermions}
 \partial_t \hat{\psi}_x = h\lrs{\vec{A}}{}_{xy} \hat{\psi}_y ,
 \quad
 \vec{j}_x = \vev{\hat{\psi}^{\dag}_y \frac{\partial h\lrs{\vec{A}}{}_{yz}}{\partial \vec{A}_x} \hat{\psi}_z} ,
 \\
 \label{eq:csft_eqs_bosons}
 \partial_t \vec{A}_x    = \vec{E}_x ,
 \quad
 \partial_t \vec{E}_x    =
  - \nabla \times \lr{\nabla \times \vec{A}}{}_{x} - \vec{j}_x - \vec{j}^{ext}_x,
\end{eqnarray}
where $h_{xy}\lrs{\vec{A}}$ is the single-particle fermionic Hamiltonian and $\vec{j}^{ext}_x$ is the external current which creates the external electromagnetic fields $\vec{A}_x^{ext}$, $\vec{E}_x^{ext}$. The fermionic current $\vec{j}_x$ in the equation for $\vec{E}_x$ leads to the back-reaction of fermions on the electromagnetic fields, to which we refer as ``backreaction'' for the sake of brevity, i.e. ``no backreaction'' means that $\vec{j}_x$ was omitted from (\ref{eq:csft_eqs_bosons}).

In the present study $h_{xy}\lrs{\vec{A}}$ is either the well-known Wilson-Dirac Hamiltonian $h^{wd}_{xy} = -i \alpha_{i} \lrs{\nabla_{i}}{}_{xy} + \gamma_0 \Delta_{xy}/2$ with zero mass or the overlap Hamiltonian \cite{Creutz:01:1}, defined as
\begin{eqnarray}
\label{eq:Overlap_hamiltonian}
 h^{ov} = \gamma_0 + \gamma_0 \gamma_5 \sign\lr{K},
 \quad
 K = \gamma_5 \gamma_0 \lr{ h^{wd} - \rho \gamma_0 } .
\end{eqnarray}
Here  $0 < \rho < 2$, $\sign\lr{K}$ is the matrix sign function of the Hermitian kernel $K$ and we have suppressed all matrix indexes and arguments for simplicity. In both cases $v_F = 1$. The matrix sign function $\sign\lr{K}$ can be defined, for example, in terms of the eigenstates $\ket{\phi_n}$ and eigenvalues $\lambda_n$ of $K$ as $\sign\lr{K} = \sum\limits_n \sign\lr{\lambda_n} \ket{\phi_n} \bra{\phi_n}$. In order to calculate the derivative of the single-particle Hamiltonian over the gauge field $\frac{\partial h}{\partial \vec{A}_x}$, which enters the fermionic current, we use the expression analogous to the first-order perturbative correction in quantum mechanics: ${\frac{\partial}{\partial \vec{A}_x} \ket{\phi_n} = \sum\limits_{m \neq n} \frac{\ket{\phi_m} \bra{\phi_m} \frac{\partial}{\partial \vec{A}_x} K \ket{\phi_n}}{\lambda_n - \lambda_m}}$. Since the kernel $K$ is expressed in terms of a local, sparse operator $h^{wd}$, the derivative $\frac{\partial}{\partial \vec{A}_x} K$ can be explicitly calculated. The expression for $\frac{\partial}{\partial \vec{A}_x} K$ involves also the contribution proportional to $\frac{\partial}{\partial \vec{A}_x} \sign\lr{\lambda_n}$ which only becomes nonzero when one of the eigenvalues $\lambda_n$ crosses zero. However, in practice we never observed such crossings, and hence we neglected this term. We finally obtain
\begin{eqnarray}
\label{eq:Overlap_vector_current}
 \frac{\partial h^{ov}}{\partial \vec{A}_x} = \sum\limits_{n \neq m}
 \lr{\frac{\sign\lr{\lambda_n} - \sign{\lambda_m}}{\lambda_n - \lambda_m}} \,
 \gamma_0 \gamma_5 \ket{\phi_n} \bra{\phi_n} \gamma_5 \gamma_0 \frac{\partial h^{wd}}{\partial \vec{A}_x} \ket{\phi_m} \bra{\phi_m} .
\end{eqnarray}

In this exploratory study on small lattices, we have obtained the eigenstates $\ket{\phi_n}$ and the eigenvalues $\lambda_n$ numerically using \texttt{LAPACK}, and calculated the current using the explicit expression (\ref{eq:Overlap_vector_current}). This brute force approach results in a $V^4$ scaling of the CPU time, where $V$ is the lattice volume, which has to be contrasted with the $V^2$ scaling for simulations with Wilson-Dirac Hamiltonian. Approximately half of CPU time is spent on the calculation of eigensystem of $K$ using \texttt{LAPACK}, and another half is taken by the matrix-vector multiplications and summation over $n$ and $m$ in (\ref{eq:Overlap_vector_current}). In order to get rid of the $V^4$ scaling which becomes prohibitively expensive at large lattices, one could use e.g. the minmax polynomial approximation for the sign function, as is common in Monte-Carlo simulations with overlap fermions. Work in this direction is in progress.

\section{Comparison of numerical results from Wilson-Dirac and overlap fermions}
\label{sec:Numerical_results}

In this Section we compare the analytic solutions obtained in Section~\ref{sec:CMW_in_magnetic_field} with the results of classical-statistical simulations with Wilson-Dirac and overlap fermionic Hamiltonians $h^{wd}$ and $h^{ov}$ defined in the previous Section~\ref{sec:CSFT}. We work on lattices of size $L \times L \times L$ with $L =200$ for Wilson-Dirac fermions and $L = 25$ for overlap fermions with periodic boundary conditions for gauge fields and fermions, and introduce a constant magnetic field with flux $\Phi = L$ through the $x_1 x_2$ plane, which corresponds to magnetic field strength $B = \frac{2 \pi \Phi}{L^2} = \frac{2 \pi}{L}$. The reason for introducing magnetic flux which is a multiple of lattice size is that in this case the magnetic translation group \cite{Wiese:08:1} is equivalent to the group of lattice translations, and translational invariance in the $x_1 x_2$ plane is not broken by the magnetic field. Thus we can consistently assume that electromagnetic fields are homogeneous. The fermions are assumed to be initially in the equilibrium state at zero temperature and density.

For Wilson-Dirac fermions, there is no unique definition of axial charge density, and we use the simplest one: $q^{wd}_A = \vev{\hat{\psi}^{\dag} \gamma_5 \hat{\psi}}$. For overlap fermions, the conserved axial charge can be defined as \cite{Creutz:01:1}
\begin{eqnarray}
\label{eq:Overlap_chiral_charge}
 q^{ov}_A = \vev{\hat{\psi}^{\dag} \gamma_5 \lr{1 - \gamma_0 h^{ov}/2} \hat{\psi}} .
\end{eqnarray}

\begin{figure}[h!tpb]
\centering
 \includegraphics[width=0.48\linewidth,angle=0]{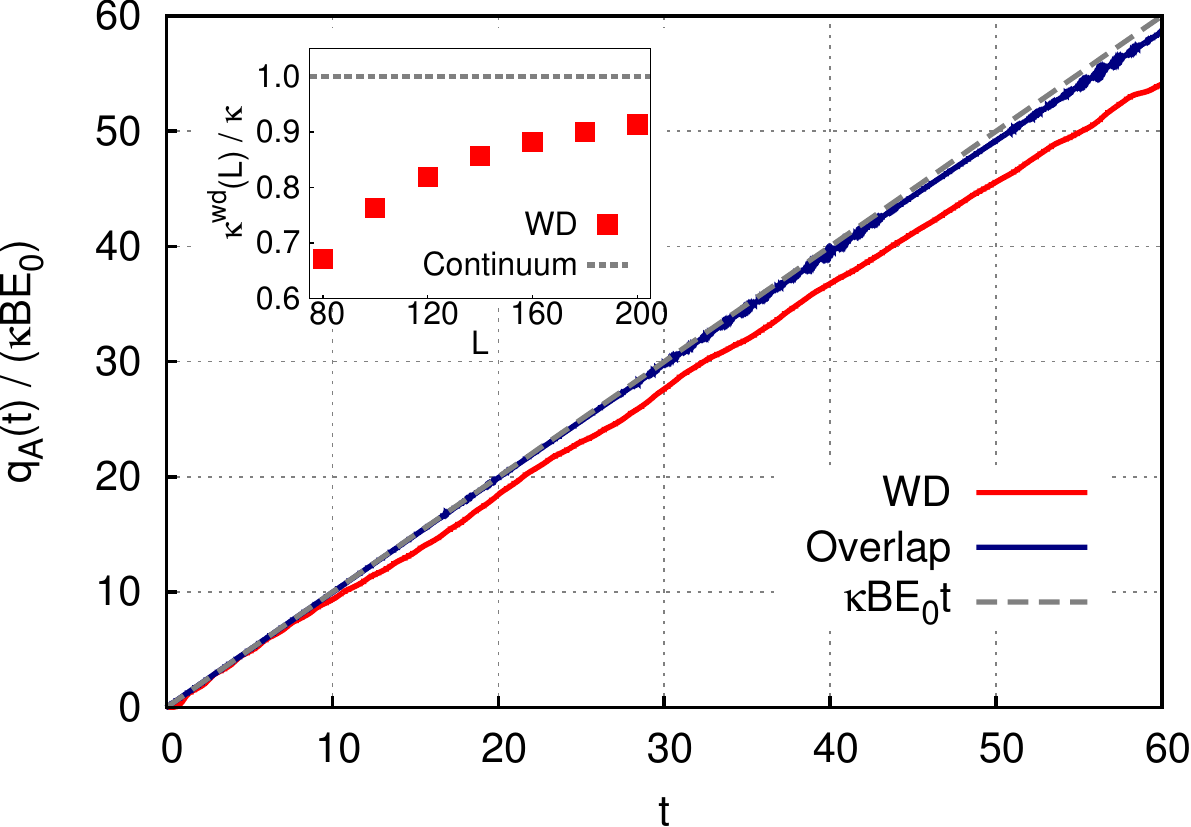}
 \includegraphics[width=0.48\linewidth,angle=0]{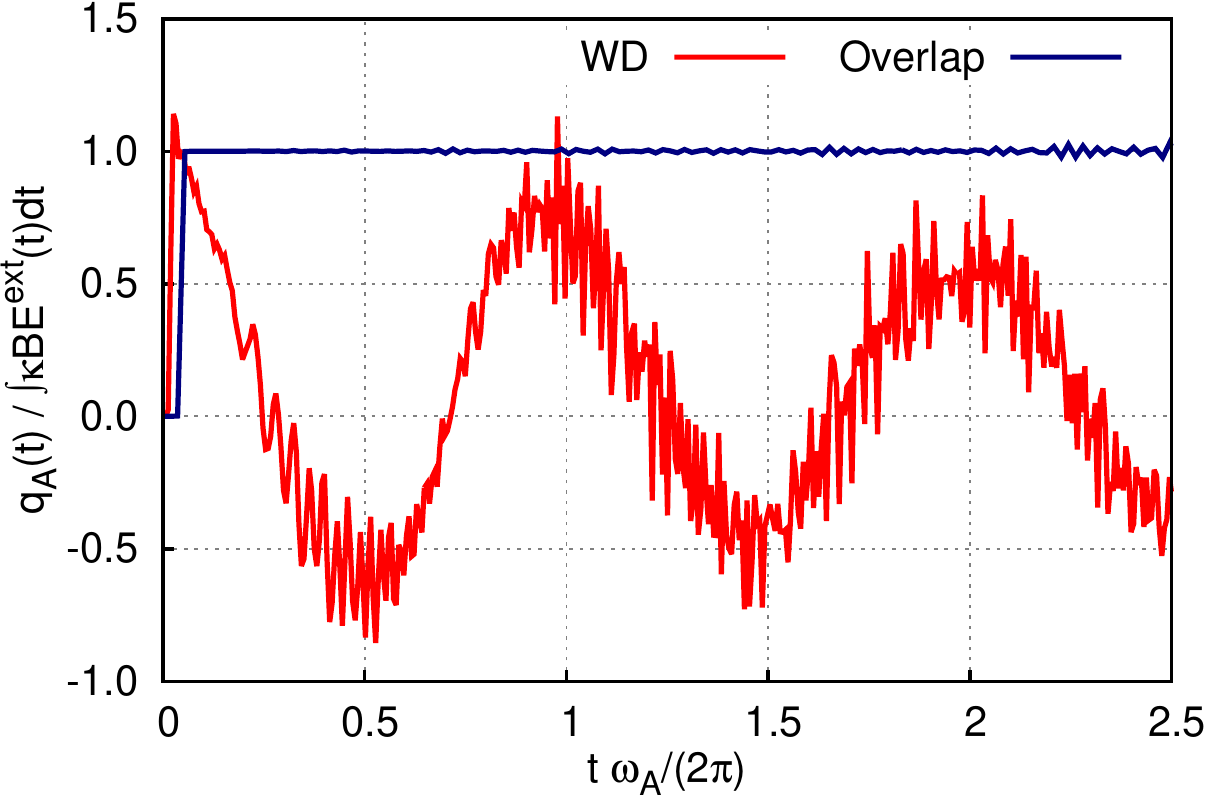}\\
 \caption{Time dependence of axial charge density in parallel external magnetic and electric field $E^{ext}(t) = E_0$ (on the left) and $ E^{ext}(t) = E_0 \exp\left( -\frac{(t-t_0)^2}{2\tau^2} \right) $ with $t_0=2.5$ and $\tau \omega_A = 2 \cdot 10^{-2}$ (on the right) for Wilson-Dirac (red line) and overlap (blue line) fermions without backreaction, compared with the continuum anomaly result $q_A = \int^t \kappa B E^{ext}(t^\prime) dt^\prime$. The inset on the left plot shows the lattice size dependence of the effective anomaly coefficient $\kappa^{wd}(L)$ for Wilson-Dirac fermions.}
\label{fig:const_E_anomaly}
\end{figure}

First, we consider the case of homogeneous and constant external electric field $E^{ext}\lr{t = 0} = E_0$ and disregard the effects of backreaction. On Fig.~\ref{fig:const_E_anomaly} we compare the time dependence of the axial charge density for Wilson-Dirac (red line) and overlap (blue line) fermions with the expected linear growth of the axial charge density $q_A = \kappa E_0 B \, t$ (dashed gray line). We see that while overlap fermions are simulated on the significantly smaller lattice, they still reproduce the continuum anomaly with a much better precision than Wilson-Dirac fermions. The inset of Fig.~\ref{fig:const_E_anomaly} shows the lattice size dependence of the effective anomaly coefficient $\kappa^{wd}\lr{L}$ which relates $E_0 B t$ and $q_A\lr{t}$ for Wilson-Dirac fermions. One can see that the approach to the continuum result is rather slow, so that even on the lattices as large as $200 \times 200 \times 200$ with magnetic field $B = \frac{2 \pi}{L} = 0.0314$ lattice artifacts result in $\sim 10 \%$ corrections to the anomaly.

\begin{figure}[h!tpb]
\centering
 \includegraphics[width=0.48\linewidth,angle=0]{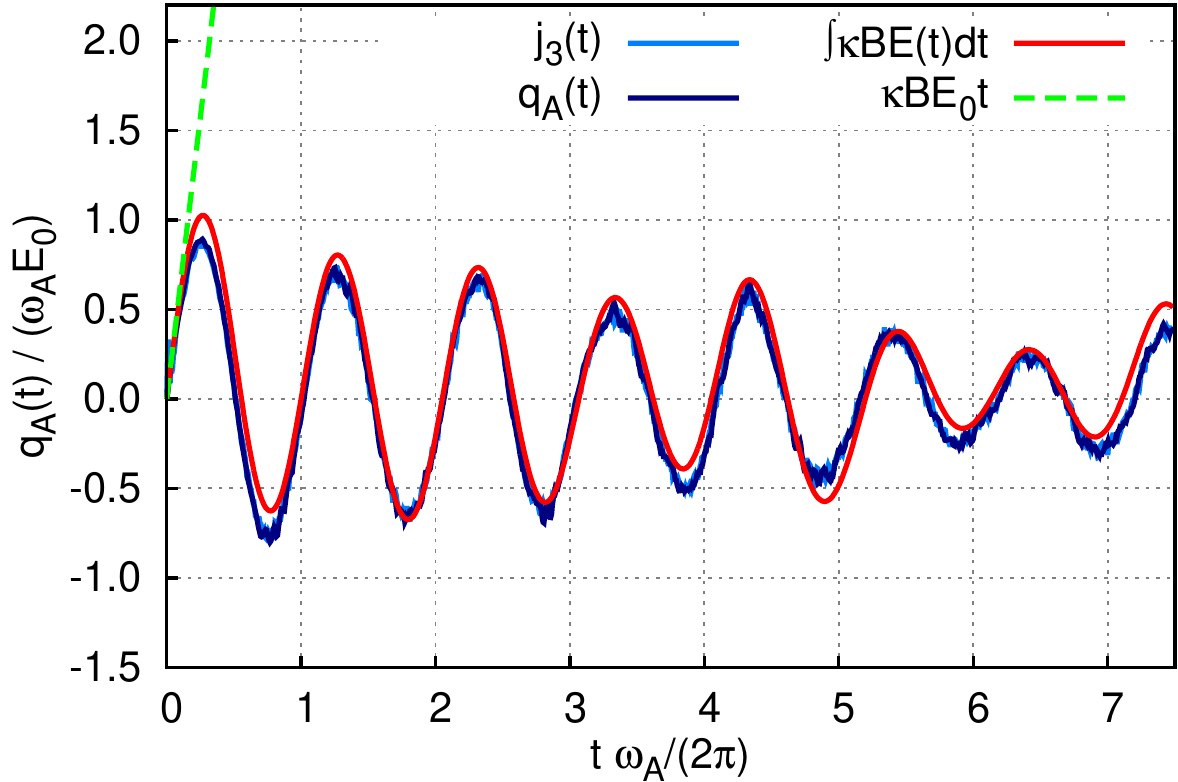}
 \includegraphics[width=0.48\linewidth,angle=0]{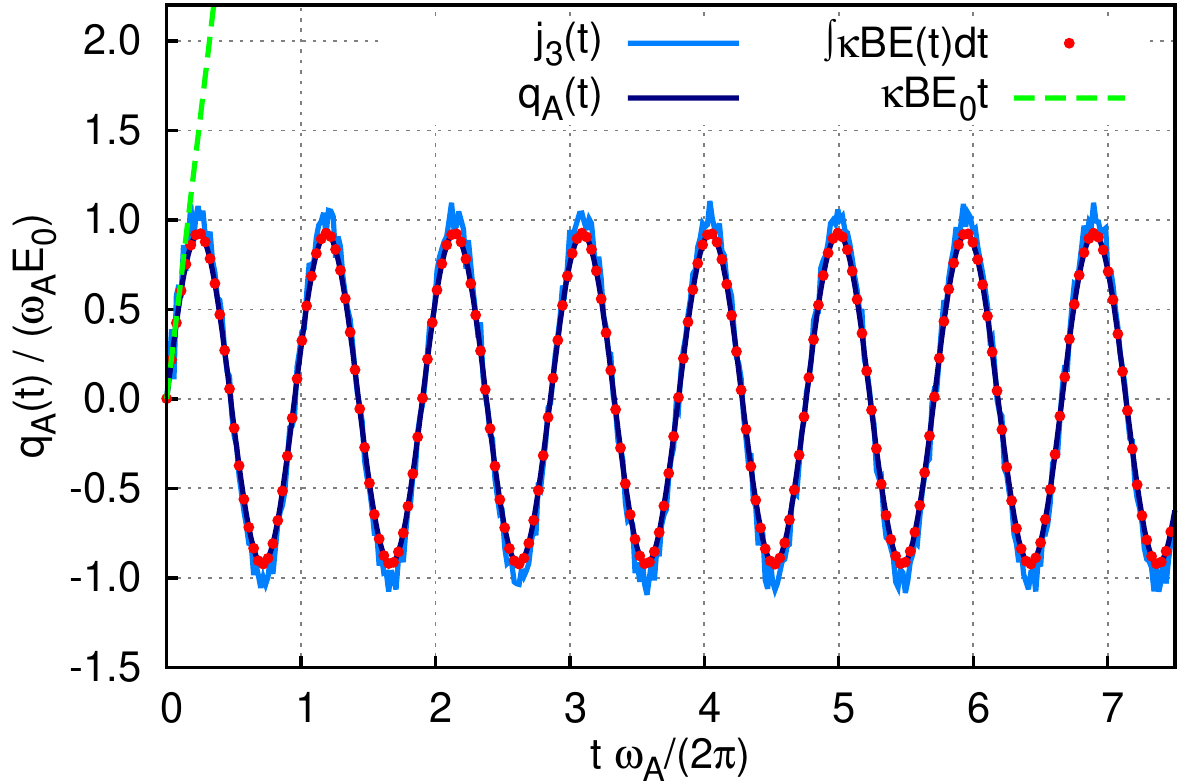}\\
 \caption{A comparison of axial charge density $q_A(t)$ (light blue line) and current density $j_3\lr{t}$ (dark blue line) with the time-integrated gauge anomaly $\kappa \int\limits_0^t d\tau \vec{E} \cdot \vec{B}$ (red line) in the presence of backreaction. On the left: for Wilson-Dirac fermions, on the right: for overlap fermions.}
\label{fig:const_electric_field}
\end{figure}

Next, we consider the effect of back-reaction of fermions on the gauge fields in the case when the external field $E_0$ is still constant in time. On Fig.~\ref{fig:const_electric_field} we compare the time dependence of the expectation values of the axial charge density and electric current for Wilson-Dirac (on the left) and overlap (on the right) fermions with the integral $\kappa \int\limits_0^t d\tau \vec{E} \cdot \vec{B}$. For both types of fermions we see the expected oscillations with the frequency $\omega_A$. For Wilson-Dirac fermions these oscillations seem to slowly decay, showing interesting similarity with the results of holographic calculations of \cite{Ammon:16:1}. However, oscillations do not decay completely and start growing after some moment of time, suggesting that this could be an artifact of Wilson term.

For overlap fermions the fermionic and the gauge parts of the anomaly equation coincide with a very good precision (of order of $10^{-8}$), but for Wilson fermions there is a visible difference. On the other hand, for Wilson-Dirac fermions the axial charge density and the electric current coincide with much better accuracy than for overlap fermions. This could be probably attributed to a much smaller lattice size for overlap fermions, and hence larger corrections due to distortions of higher Landau levels.

\begin{figure}[h!tpb]
\centering
 \includegraphics[width=0.48\linewidth,angle=0]{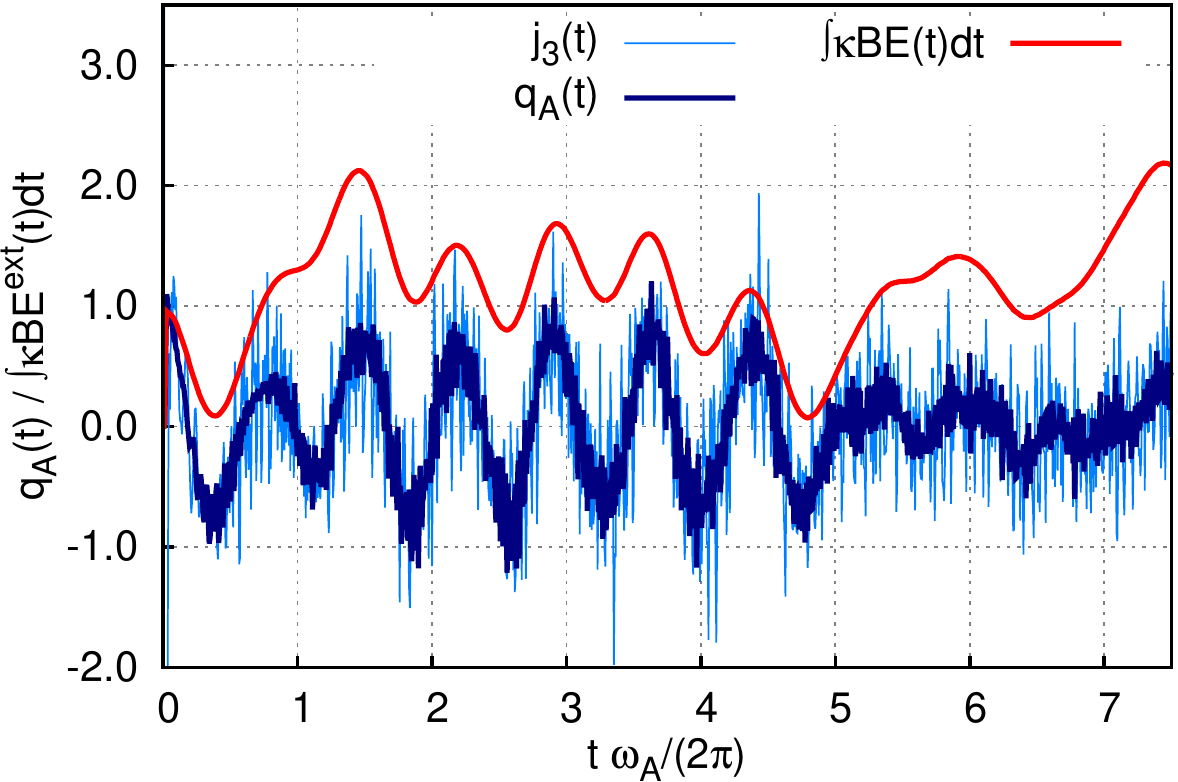}
 \includegraphics[width=0.48\linewidth,angle=0]{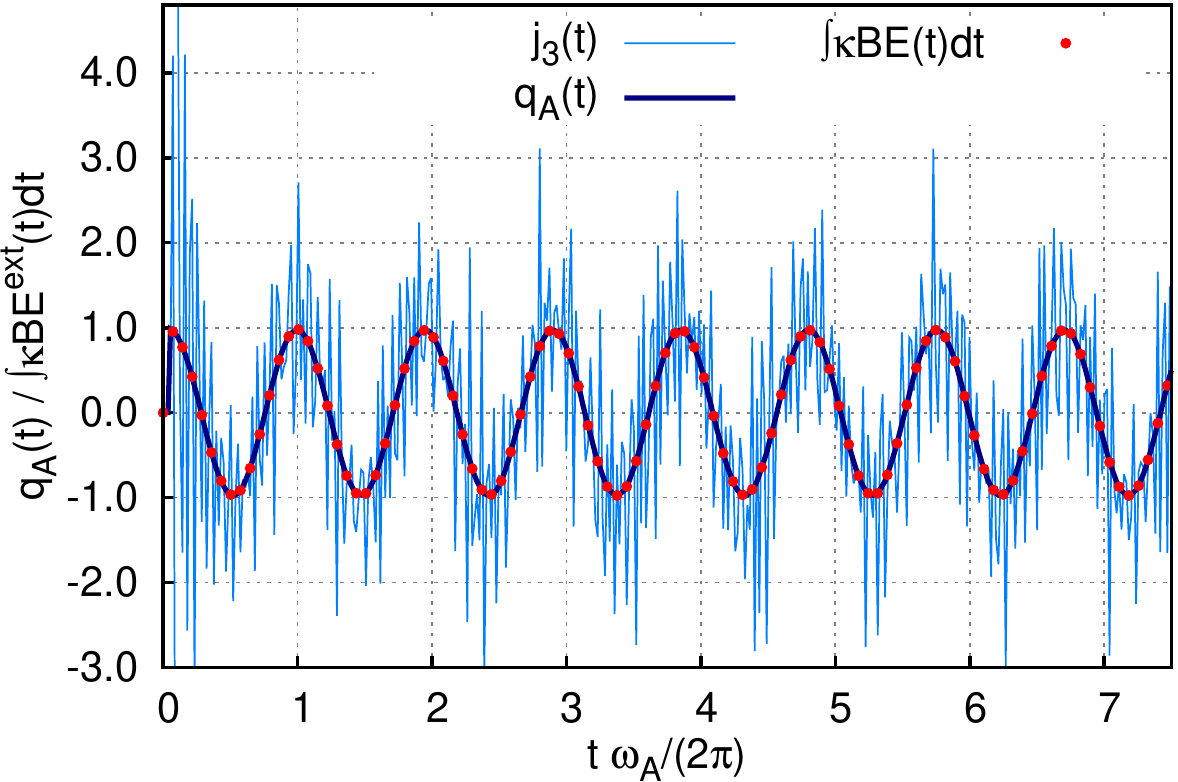}\\
 \caption{Time dependence of the axial charge density $q_A(t)$, electric current density $j_3(t)$ and the time-integrate gauge anomaly $\kappa \int dt B E(t)$ after a short pulse of external electric field for Wilson-Dirac (on the left) and overlap (on the right) fermions in the presence of backreaction.}
\label{fig:anomaly_after_short_pulse}
\end{figure}

Finally, we study the case of a short Gaussian-shaped pulse of external electric field
$ E^{ext}(t) = E_0 \exp\left( -\frac{(t-t_0)^2}{2\tau^2} \right) $, with $t_0 = 2.5$ and $\tau \omega_A = 2 \cdot 10^{-2} \ll 1$, where the role of higher Landau levels should be significant. If backreaction is disregarded, after the external electric field is switched off the axial charge of overlap fermions stays at a perfectly constant value consistent with the anomaly equation, as illustrated on the right plot on Fig.~\ref{fig:const_E_anomaly}. Remembering the relation between the axial charge density and the electric current along the magnetic field, we can regard this situation as a real-time manifestation of the Chiral Magnetic Effect. Namely, with a short pulse of electric field parallel to magnetic field we have created nonzero axial charge in the system, and it results in non-dissipative electric current flowing along the magnetic field even when electric field becomes zero. However, this situation will be most likely unstable for sufficiently large volumes due to the chiral plasma instability phenomenon \cite{Yamamoto:13:1,Buividovich:15:2} once we allow for fully 3D electromagnetic fields. In this case any small perturbation would result in quickly growing transverse helical electromagnetic fields and subsequent decrease of axial charge density and electric current.

In contrast, for the case of Wilson-Dirac fermions without backreaction the axial charge is not conserved at all after the same short pulse, as illustrated on the right plot of Fig.~\ref{fig:const_E_anomaly}. Instead we observe nearly periodic, slightly decaying oscillations of axial charge density, with some irregular short-range fluctuations on top. These oscillations originate in explicit chiral symmetry breaking due to the Wilson term, which becomes significant for higher energy levels which are excited by the short pulse of electric field. The slow decay of $q_A\lr{t}$ is similar to the one observed for constant external electric field with backreaction (left plot on Fig.~\ref{fig:const_electric_field}), which again suggests that it might be an artifact of Wilson-Dirac fermions.

On Fig.~\ref{fig:anomaly_after_short_pulse} we illustrate the effect of backreaction on the time dependence of the chiral charge, electric current and the time-integrated gauge anomaly $\kappa \int dt B E(t)$ for Wilson-Dirac (on the left) and overlap (on the right) fermions subject to the same short pulse of external electric field. For overlap fermions we still observe regular oscillations with frequency $\omega_A$. The fermionic and the gauge contributions to the anomaly equation agree very well, but electric current $j_3\lr{t}$ exhibits significant short-range oscillations. The origin of this oscillations is again the dynamics associated with higher Landau levels. With Wilson-Dirac fermions, however, the gauge and the fermionic contributions to the anomaly disagree significantly already at early times, and are completely de-correlated at late times $t \omega_A/\lr{2 \pi} \gtrsim 6$. The electric current and the axial charge still agree quite well up to quite large short-scale fluctuations. By performing the Fourier transform of numerical data for $q_A\lr{t}$ we have found that for Wilson-Dirac fermions we have a mixture of two oscillations with frequencies approximately equal to $0.6 \omega_A$ and $1.5 \omega_A$.

%\bibliographystyle{mybibstyle}
%\bibliography{Buividovich}

\end{document}